\documentstyle[aps,pra,epsfig,amstext,amsmath,eqsecnum]{revtex}

\title{Simulation of Markovian quantum dynamics on quantum logic networks}
\author{M. Koniorczyk$^{1,2}$, V. Bu\v zek $^{3,4}$, P. Adam$^{1,2}$ and
  A. L\' aszl\' o$^5$}
\address{
$^1$Department of Nonlinear and Quantum Optics,
    Research Institute for Solid State Physics and Optics,\\
    Hungarian Academy of Sciences,
    P.O. Box 49, H-1525 Budapest, Hungary
\\$^2$Institute of Physics, 
    University of P\'ecs, 
    Ifj\'us\'ag \'ut 6. H-7624 P\'ecs, Hungary 
\\$^3$Research Center for Quantum Information, 
    Slovak Academy of Sciences
    D\'ubravsk\'a cesta 9,
    842 28 Bratislava, Slovakia
\\$^4$Faculty of Informatics, Masaryk University,
Botanick\' a 68 a, Brno 602 00, Czech Republic
\\$^5$Institute of Mathematics, 
    University of P\'ecs, 
    Ifj\'us\'ag \'ut 6. H-7624 P\'ecs, Hungary 
}
\date{\today}

\begin{document}
\draft
\maketitle

\newcommand{\tr}{\mathop{\mbox{tr}}\nolimits}
\newcommand{\rank}{\mathop{\mbox{rank}}\nolimits}

\begin{abstract}
We study how generators of Markovian dynamics of a
  qubit can be simulated using a programmable quantum processor.
\end{abstract}

\maketitle

\section{Introduction}
\label{sect:intro}

Quantum computing offers a new perspective in simulating physical
systems\cite{manin,intjtheor21_467}. While the simulation of a larger
quantum system is intractable for a classical computer, it should be
possible of course on a quantum computer. Quantum computers are, from
a theoretical point of view, highly simplified quantum systems. A
simulation of a real physical system on such an arrangement is
therefore also interesting necessarily because of the usefulness of
the simulation result, but also because it may reveal fundamental
aspects of the real physical system, which may be obscure otherwise.

Maybe the most general problem of this kind would be the simulation of
general quantum operations\cite{Krauss}, the most general dynamics a
quantum system can undergo. The idea is to represent the initial state
of the system on a quantum bit array, and supplement it with another
array embodying the environment. Subjecting this system to the effect
of a quantum logic network, and dropping the environment bits, the
resulting state of the data should be obtained. Naturally, having
infinite resources, i.e. arbitrary number of ``environment'' bits, any
operation can be carried out by realizing a unitary representation of
the process. For a system with a $k$ dimensional Hilbert space, this
requires a $k^2$ dimensional environment.  Lloyd
conjectured~\cite{science273_1073}, that a general quantum operation on
a system with a $k$ dimensional Hilbert-space can be simulated with a
$k$-dimensional environment, which may be in a mixed sate initially.
This conjecture was falsified by Terhal et al.~\cite{pra60_881}, who
give bounds to the dimensionality of the environment required to
simulate certain quantum operations on a single qubit system. Zalka
and Rieffel\cite{jmathphys43_4376} give a simpler proof via explicit
construction of counterexamples.  More optimal possibilities for the
simulation, such as closed loop schemes were discussed in detail in
Refs.~\cite{prl83_4888,pra65_010101}.

Programmable quantum gate arrays or quantum
processors~\cite{prl79_321,quantph0102037,fortschr49_987,pra65_022301}
provide another possible approach to the problem.  In this case, the
parameters of the dynamics should be stored in the initial state of
the ``environment'' (which play a role of the ``program register'' -
see below).  The quantum processor has two input registers, one
storing the state of the system to be simulated (the ``data
register''), while the other contains the data of the operation to be
performed.  After the operation of the circuit, the remains of the
program state in the program register may either be dropped or be
subjected to a measurement. The latter is the probabilistic regime of
the quantum processor. The result is accepted, if the measurement
yields a given result. It was shown by Hillery et
al.~\cite{fortschr49_987,pra65_022301}, that in this case, any
operation can be implemented, though some of them with quite low
probability of success. Recently Vidal et al.~\cite{prl88_047905} have
presented a probabilistic scheme implementing unitary transformations
with rather high probability.

In this paper we restrict ourselves to the ``deterministic'' case,
that is, the remains of the program register are dropped.  Hillery et
al.~\cite{pra66_042302} have studied the possibility of implementing
general quantum maps in such a scheme. They have shown, that there are
certain limitations, e.g., an amplitude damping channel cannot be
implemented. In addition to this, we show that different quantum
circuits such as quantum teleportation~\cite{bennett} can be used as a
programmable quantum circuit for some purpose, as it is related to a
phase damping channel~\cite{telepdecoh}.

We focus here on a specific subset of quantum operations, namely
Markovian dynamics. These processes are the most important ones in the
description of quantum decoherence. The question of simulating
Markovian dynamics in the quantum computing context was investigated
by Bacon et al.~\cite{pra64_062302}. As one of their main results,
these authors provide a decomposition rule to build more complex
dynamics from simpler ``primitive'' operations. On the other hand,
Bacon et al.  apply the formalism of Gorini, Kossakowski and Sudarshan
(GKS)~\cite{jmathphys17_821}, providing an excellent framework for the
treatment of Markovian processes.  We approach the problem of
simulating Markovian dynamics differently in a way that is related to
programmable quantum gate arrays. We restrict ourselves to single
qubit dynamics, with a single qubit program register. We intend to
simulate the \emph{generator} of Markovian dynamics on a programmable
quantum logic array, as the generator characterizes the time evolution
completely in this case.  A single run of the processor simulates an
\emph{infinitesimal} time step.  As a further restriction we require
that the length \emph{infinitesimal time step} should be encoded into
the initial state of the program register.

The paper is organized as follows: in Section \ref{sect:markov} we
review some definitions concerning Markovian dynamics, introducing
Lindbladians, the generators of the dynamics. The standard GKS
matrix-representation introduced by Gorini, Kossakowski and Shudarsan
in Ref.~\cite{jmathphys17_821} is reviewed, and a practical recipe is
given for finding the GKS matrix from of a Lindbladian operator in the
one-qubit case, utilizing the affine representation. In
Section~\ref{sect:lsim} we describe the general idea of simulating a
generator, as understood in this paper. Section~\ref{sec:progu}
describes a reversible scheme capable of simulating a phase damping
channel about an arbitrary axis. A geometrical interpretation of the
result is also given. In Section~\ref{sect:telepcont} we discuss an
application of Bennett's teleportation scheme in this context. In
section \ref{sect:summary} the results are summarized and conclusions
are drawn.

\section{Markovian dynamics of a qubit}
\label{sect:markov}

In this Section we review the definition of Markovian semigroup and
its generators very briefly.  The most general operation, that a state
of a qubit can undergo is described by a completely positive linear,
trace preserving map acting on the set of density operators, also
called a superoperator. This may be written in the
Krauss-representation\cite{Krauss} as
\begin{equation}
  \label{eq:Krauss}
  {\cal{E}} (\varrho)=\sum\limits_k E_k\varrho E_k^\dag
\end{equation}
where the $E_k$-s are positive operators such that $\sum
E_k^\dag E_k=1$.

In order to introduce Markovian processes, one equips the set of
superoperators with a continuous parameter $t$ which is  called the
time. Stationary and Markovian dynamics obey the property
\begin{equation}
  \label{eq:Markov_def}
  {\cal{E}} _{t_1} {\cal{E}} _{t_2} = {\cal{E}}_{t_1+t_2},
\end{equation}
with $t_1, t_2 > 0$.
The set of superoperators with property (\ref{eq:Markov_def}) form a
one-parameter semigroup, the Markovian semigroup. We also require the
property
\begin{equation}
  \label{eq:t0}
  {\cal{E}} _{0}=\hat 1,
\end{equation}
(where $\hat 1$ stands for the identity superoperator) to hold.

The property in Eq.~(\ref{eq:Markov_def}) enables us to define the
generators of the semigroup as
\begin{equation}
  \label{eq:geners_def}
  \hat{L}(\varrho)
=\lim\limits_{t\to 0+0} \frac{{\cal{E}} _t(\varrho)-\varrho}{t}.
\end{equation}
The operator $\hat{L}$ is the infinitesimal generator of the time evolution:
\begin{equation}
  \label{eq:L1}
  {\cal{E}}_t=\exp(\hat{L}t)
=\lim\limits_{n\to \infty} \left( \hat 1 -\frac{t}{n}
    \hat{L}\right)^{-n},
\end{equation}
from which
follows, that
\begin{equation}
  \label{eq:Master}
  \frac{\partial \varrho(t)}{\partial t}=\hat{L}[\varrho(t)]
\end{equation}
known as the master equation.

Let us consider an operator $\hat L$ acting on the Hilbert space of a
qubit. The question arises, under what condition can this operator
represent a generator of the dynamical semigroup.  The answer was
given by Lindblad~\cite{lindbl}, and by Gorini, Kossakowski and
Sudarshan (GKS) \cite{jmathphys17_821}. We use the notation of the
latter authors.  According to this, the most general form of a
generator of a Markovian semigroup reads
\begin{equation}
  \label{eq:GKS}
    {\hat L}(\varrho) =
    -i [\hat H, \varrho]+
     \frac12 \sum\limits_{i,j=1}^3 C_{i,j}
                     ([\hat \sigma_i \varrho, \hat \sigma_j]
                     +[\hat \sigma_i, \varrho \hat \sigma_j]),
\end{equation}
where the $\hat{\sigma} $-s are the Pauli-matrices. The first
contribution on the right hand size describes a possible unitary
evolution, the Hamiltonian $\hat H$ being a Hermitian matrix which can
be chosen to be traceless without the loss of generality. The second
contribution describes the stochastic part of the evolution. The
Hermitian positive semidefinite matrix $C_{i,j}$ is called the GKS
matrix, and it contains all the information on the nature of the
dynamics. Note that in order to preserve the trace of the density
matrix, $\hat L \varrho$ should be traceless. In fact,
Eq.~\eqref{eq:GKS} describes the most general linear operator taking
$\varrho$ to a traceless matrix.

In the following we describe how to extract the Hamiltonian $\hat H$
and the GKS matrix $C$, if one is given an arbitrary function of a
single-qubit density matrix $\varrho'=\hat L \varrho $ In order to do
so we utilize the relation between the GKS matrix and the affine
representation described also in ~\cite{pra64_062302}.

A generic density matrix can be expanded on the basis of the three
Pauli-matrices and the unit operator, obtaining its usual real
3-vector representation $\underline r$ displayable on the unit radius
Bloch sphere:
\begin{equation}
  \label{eq:bball}
{\underline r}=
    \begin{pmatrix}
      r_1\cr r_2\cr r_3
    \end{pmatrix}
, \quad r_i =\tr (\varrho \hat \sigma_i)\; ,
\end{equation}
so that
\begin{equation}
  \label{eq:bballrev}
  \varrho= \frac{1}{2} \left( \hat 1 + \sum\limits_{i=1}^3
   r_i\hat{\sigma}_i \right)\; .
\end{equation}
Direct calculation shows that the real 3-vector $\hat L
[\underline{r}]$ corresponding to the most general $\hat L \varrho$ in
Eq.~\eqref{eq:GKS} reads
\begin{equation}
  \label{eq:gksbloch}
\hat L [\underline{r}]=
\begin{pmatrix}
0   & -h_3 & h_2 \cr
h_3 & 0    & -h_1 \cr
-h_2 & h_1  &   0
\end{pmatrix}
\underline{r}
+
\begin{pmatrix}
  -2( C_{22} +C_{33}) & C_{12}+C_{21} & C_{13}+C_{31} \cr
  C_{12}+C_{21} & -2( C_{11} +C_{33}) & C_{23}+C_{32} \cr
  C_{13}+C_{31} & C_{23}+C_{32} & -2( C_{11} +C_{22})
\end{pmatrix}
\underline{r}
+
2i
\begin{pmatrix}
  C_{23} -C_{32} \cr C_{31} -C_{13} \cr C_{12} -C_{21}
\end{pmatrix}
,
\end{equation}
where
\begin{equation}
  \label{eq:Hbas}
  \hat H = \frac12 \sum\limits_{i=1}^3 h_i \hat \sigma_i.
\end{equation}
Thus $\hat L$ appears as an affine linear mapping in the 3 space, the
above mapping is called the affine representation of $\hat L$.

If we are now given an arbitrary function $\varrho'$, which is
traceless and depends on the components of the arbitrary density
matrix $\varrho$ linearly, we can find the corresponding
$\underline{r'}$ according to Eq.~\eqref{eq:bball} as a function of
the components of $\underline{r'}$ representing $\varrho$. This is a
linear inhomogenous operator, which can be always decomposed into a
sum of a homogenous antisymmetric operator, a homogenous symmetric
operator and a vector representing the inhomogenity. For qubits, this
decomposition of the affine representation of the generator is quite
meaningful: according to Eq.~\eqref{eq:gksbloch}, the information on
the unitary part of the generator, i.e. the Hamiltonian is encoded
into the antisymmetric part of this operator, while the real and
imaginary parts can be found from the symmetric part and the
inhomogenity respectively.

In the absence of the inhomogenity the generator is zero for the
identity operator:
\begin{equation}
  \label{eq:unitpres}
  \hat L \hat I=0,
\end{equation}
thus it describes a unital evolution, which preserves the identity
operator. The inhomogenity appears in the complex part of the elements
of the GKS matrix: for qubits, real GKS matrices correspond to unital
dynamics.

Thus equipped with Eq.~\eqref{eq:gksbloch}, we have the recipe how to
find the standard GKS form of a generator of a dynamical semigroup for
a quantum bit.

\section{Simulation of infinitesimal generators}
\label{sect:lsim}

We intend to simulate the infinitesimal generator $L\hat \varrho$ of
Markovian dynamics, on a single quantum bit. This can be understood in
several ways. As an elementary step we consider the application of a
quantum processor: an arrangement of quantum logic gates, and possibly
measurements, acting on the qubit in argument, and certain ancillary
systems. The ancillary systems can be used as a ``program register'':
their state can influence the action of the processor on the qubit in
argument, which constitutes a ``data bit'' in this context.

The next question might be, how to interpret the time. A possible generic
approach would be to regard the single run of the processor as a
\emph{finite} time step $\Delta t$. The repeated application of the
processor on the data bit results then in a discrete time evolution.
One can then examine if this is a stroboscopic version of some valid
continuous time evolution, and search for the proper master equation.
This approach will be discussed elsewhere.  Here we adopt a simpler
interpretation: we expect a single run of processor to simulate an
\emph{infinitesimal} time step:
\begin{equation}
  \label{eq:cpuwork}
  \varrho_{\text{out}}=\varrho_{\text{in}}+\hat L \varrho_{\text{in}}
  dt +{\mathcal{O}}(dt^2),
\end{equation}
where $dt$ should be encoded in the $|\Psi_{\text{prog}}\rangle$ of
the program register.  The entire evolution can then be approximated
with some accuracy by running this process many many times. This
implements the equidistant first order Euler method\cite{nr} of
solving Eq.~(\ref{eq:Master}), but time step, and thereby the
\emph{accuracy} is encoded quantum mechanically.  Of course, the
simulation is completely accurate if $dt\to 0$, and the number of
repetitions tends to infinity.  We remark here, that simulation of
decoherence mechanisms with an array of quantum gates has proven to be
fruitful in other problems too~\cite{prl88_097905,pra65_042105}.

Physically, the simulation scheme can be envisaged as a simple
collision model: The data qubit is represented by a physical system
localized in space.  It interacts with flying program bits represented
by e.g. spin of particles emerging from an oven. Each program bit
causes the system to evolve a small time step further.

It follows from Eq.~(\ref{eq:cpuwork}), that there must exist a
program state, for which $dt=0$, and thus
$\varrho_{\text{out}}=\varrho_{\text{in}}$, in the quantum processor
terminology we would say the processor implements the unit operator.
It is a natural requirement for this kind of semigroup simulation.
Thus our scheme is to some extent similar to the idea of simulating a
reservoir with beam-splitters of transmittance around unity in quantum
optics~\cite{pra52_2401} or a quantum homogenization processes as
described in Ref.~\cite{pra65_042105}.

\section{A scheme without a measurement}
\label{sec:progu}

\begin{figure}[htbp]
  \begin{center}
    \includegraphics[width=0.4\textwidth]{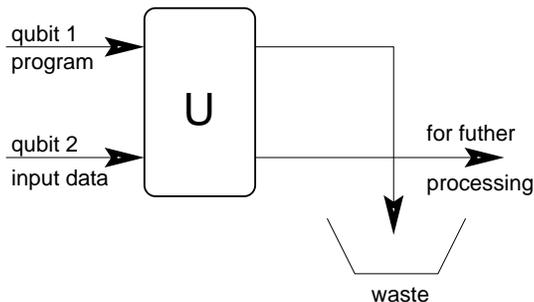}
    \caption{The quantum network for the deterministic scheme. A single run of the network simulates an infinitesimal time-step in the evolution.}
    \label{fig:scheme}
  \end{center}
\end{figure}

First we consider a simple deterministic scheme with a single
ancillary qubit, as depicted in Fig.~\ref{fig:scheme}.
The first bit contains the program, which is most generally
\begin{equation}
  \label{eq:psiprog}
  |\Psi_{\text{prog}}\rangle=\sqrt{1-\varepsilon} e^{i\chi}|0\rangle +
   \sqrt{\varepsilon}|1\rangle,\quad 0\leq\varepsilon\leq 1,\ \chi\ \text{real}.
\end{equation}
We don't consider mixed program states here, we want to study how
decoherence associated with Markovian processes appears purely
quantum-mechanically, and a mixed program state would imply some
prescribed classical stochasticity.  The input data is in bit 2, in
the state $\varrho_{\text{in}}$. After the operation of the processor,
the contents of the program register are dropped, and bit 2 contains
the output
\begin{equation}
  \label{eq:oper}
 \varrho_{\text{out}}=\mathop{\mbox{Tr}}\nolimits_1
\left( \hat U ( |\Psi_{\text{prog}}\rangle \langle \Psi_{\text{prog}}|
\otimes \varrho_{\text{in}}) U^\dag \right)
\end{equation}
which is passed for further processing.

To have the identity operator implemented, there should be a program
state for which $\varrho_{\text{out}}=\varrho_{\text{in}}$ holds. We
chose this to be the program state $|0\rangle$.  The most general
processor that is possible under such circumstances is a controlled U gate:
\begin{equation}
  \label{eq:cpu}
  \hat U=
  \begin{pmatrix}
    \hat 1 & 0 \cr 0 & \hat U_2
\end{pmatrix}
\end{equation}
The lower right corner is an $SU(2)$ block:
\begin{equation}
  \label{eq:su2}
\hat U_2=
\begin{pmatrix}
 \cos (\frac{\theta }{2}) e^{-i\frac{\phi+\psi}{2}} &
                    -\sin (\frac{\theta }{2})  e^{-i\frac{\phi-\psi}{2}} \cr
                   \sin (\frac{\theta }{2}) e^{i\frac{\phi-\psi}{2}}&
                   \cos (\frac{\theta }{2}) e^{i\frac{\phi+\psi}{2}}
\end{pmatrix},
\end{equation}
where we use the standard Euler angle parametrization in the $y$
convention~\cite{yconv}.

Now we can evaluate Eq.~(\ref{eq:cpuwork}) with a generic input data
state of Eq.~(\ref{eq:bballrev}), and program of
Eq.~(\ref{eq:psiprog}), to obtain the effect of a single run of the
processor. Note that due to the orthogonality of the program states
corresponding to $\hat 1$ and $\hat U_2$ (which is a consequence of
the unitarity of $\hat U$), Eq.~\eqref{eq:oper} simplifies to
\begin{equation}
   \label{eq:rhoout1}
   \varrho_{\text{out}}=
   (1-\varepsilon) \varrho_{\text{in}}+
   \varepsilon \hat U_2 \varrho_{\text{in}} \hat U_2^\dag.
 \end{equation}
 Thus from the point of view of the effect on the input state, the
 quantum circuit does nothing else than apply the unitary operation
 $\hat U_2$ on the input state, with the probability $\varepsilon$. The
 difference is that if the operation is realized by the quantum
 circuit, then the information which disappears from the system will
 be stored in the dropped quantum state of the program register. In
 the other case the information will be entirely classical, one bit
 per infinitesimal timestep: we can be aware, whether the operation
 $\hat U_2$ was carried out or not. Note that in both of the cases the
 process is reversible, provided that we have the appropriate
 information at hand.

 According to Eq.~\eqref{eq:rhoout1}, we can write
 \begin{equation}
   \label{eq:gen1a}
   \varrho_{\text{out}}=
   \varrho_{\text{in}}+
   (\hat U_2 \varrho_{\text{in}} \hat U_2^\dag-
   \varrho_{\text{in}})\varepsilon.
 \end{equation}
 Thus we identify $\varepsilon=dt$, and comparing with
 Eq.~\eqref{eq:cpuwork} we get
 \begin{equation}
   \label{eq:gen1}
   \hat L \varrho_{\text{in}}=\hat U_2 \varrho_{\text{in}} \hat U_2^\dag-
   \varrho_{\text{in}}.
 \end{equation}
 The transformation on the Bloch ball corresponding to $\hat U_2
 \varrho \hat U_2^\dag$ is a rotation of the vector $\underline{r}$
 corresponding to $\varrho$
 \begin{equation}
   \label{eq:rot}
      \underline{r}'= {\mathcal{R}}(\theta,\phi,\psi)  \underline{r},
 \end{equation}
 where ${\mathcal{R}}$ is the appropriate element of the adjoint
 representation of $SU(2)$:
 \begin{equation}
   \label{eq:conjrep}
    {\mathcal{R}}(\theta,\phi,\psi)=
   \begin{pmatrix}
     -\sin \phi \sin \psi + \cos \theta \cos \phi \cos \psi &
-\cos \theta \cos \phi \sin \psi -\sin \phi \cos \psi &
\sin \theta \cos \phi \cr
\cos \theta \sin \phi \cos \psi + \cos \phi \sin \psi &
\cos \phi \cos \psi - \cos \theta \sin \phi \sin \psi &
\sin \theta \sin \phi \cr
-\sin \theta \cos \psi & \sin \theta \sin \psi &
\cos \theta
   \end{pmatrix}\in SO(3) \; .
 \end{equation}
For the generator in Eq.~\eqref{eq:gen1} we have thus
\begin{equation}
  \label{eq:gen1bloch}
\hat L [\underline{r}]=
\left( {\mathcal{R}}(\theta,\phi,\psi)-\hat 1\right) \underline{r}.
\end{equation}
This can be compared with Eq.~\eqref{eq:gksbloch} to extract the
properties of the generator.
The transformation in Eq.~\eqref{eq:gen1bloch} is homogenous, thus the
generated dynamics is unital.

As for the coherent part of the evolution, we get the Hamiltonian
\begin{equation}
  \label{eq:H1}
  \hat H =
  \begin{pmatrix}
    \sin (\phi+\psi) \cos ^2 \frac{\theta}{2} &
    -\frac{i}{2} \sin \theta \left(e^{-i\phi}+e^{i\psi} \right)  \\
    \frac{i}{2} \sin \theta \left(e^{i\phi}+e^{-i\psi} \right) &
    -\sin (\phi+\psi) \cos ^2 \frac{\theta}{2}
  \end{pmatrix}
\end{equation}
This is zero if $\phi+\psi=(2k+1)\pi$ or $\theta=\pi+2k\pi$ holds,
which is equivalent to $\tr U_2=0$. Thus in case of traceless
unitaries, we obtain a purely stochastic evolution in the sense that
it lacks the Hamiltonian part.

Evaluating the symmetric part of Eq.~\eqref{eq:conjrep}, and comparing
with~\eqref{eq:gksbloch}, we obtain the GKS matrix
\begin{equation}
  \label{eq:gengks}
  C_{\theta,\phi,\psi}=
  \begin{pmatrix}
    \sin ^2 \frac{\theta}{2} \sin ^2 \frac{\phi-\psi}{2} &
    \frac12 \sin ^2 \frac{\theta}{2} \sin(\psi-\phi) &
    \frac14 \sin \theta (\cos \phi - \cos \psi)
    \cr
    \frac12 \sin ^2 \frac{\theta}{2} \sin(\psi-\phi) &
    \sin ^2 \frac{\theta}{2} \cos ^2 \frac{\phi-\psi}{2} &
    \frac14 \sin \theta (\sin \phi + \sin \psi)
    \cr
    \frac14 \sin \theta (\cos \phi - \cos \psi) &
    \frac14 \sin \theta (\sin \phi + \sin \psi) &
    \cos ^2 \frac{\theta}{2} \sin ^2 \frac{\phi+\psi}{2}
  \end{pmatrix}
\end{equation}

Specifically, if $\hat U_2$ is traceless because $\psi=\pi-\phi$
holds, we obtain from Eq.~\eqref{eq:gengks} the following GKS matrix:
\begin{equation}
  \label{eq:GKSpm}
    C_{\theta,\phi}=
  \begin{pmatrix}
    \sin ^2 \frac{\theta}{2} \cos ^2 \phi &
    \frac12 \sin ^2 \frac{\theta}{2} \sin(2\phi) &
    \frac12 \sin \theta \cos \phi
    \cr
    \frac12 \sin ^2 \frac{\theta}{2} \sin 2\phi &
    \sin ^2 \frac{\theta}{2} \sin ^2 \phi &
    \frac12 \sin \theta \sin \phi
    \cr
    \frac12 \sin \theta \cos \phi &
    \frac12 \sin \theta \sin \phi &
    \cos ^2 \frac{\theta}{2}
  \end{pmatrix}
\end{equation}
The matrix in Eq.~(\ref{eq:GKSpm}) can be interpreted as follows.
Consider a unitary operator ${\mathcal{U}}\in SU(2)$ acting on the qubit's
Hilbert space, and a superoperator ${\mathcal{E}}$, which describes
unitary evolution described by the GKS matrix $C$. According to
Ref.~\cite{pra64_062302} \emph{unitary conjugation} of a superoperator
${\mathcal{E}}$, that is,
\begin{equation}
   \label{eq:unitconj}
   {\mathcal{E}}'={\mathcal{U}}^\dag {\mathcal{E}} {\mathcal{U}},
 \end{equation}
 where ${\mathcal{U}}(\varrho)={\mathcal{U}} \varrho {\mathcal{U}}
 ^\dag$ yields another superoperator describing Markovian dynamics as
 well.  The resulting GKS matrix is
 \begin{equation}
   \label{eq:gksconj}
   C'={\mathcal{R}} C {\mathcal{R}}^T,
 \end{equation}
where ${\mathcal{R}}$ is the element of $SO(3)$, the adjoint
representation of $SU(2)$ corresponding to $\mathcal{U}$,
and the $T$ stands for transposition. Thus
$\mathcal{R}$ is a real 3-rotation, which can be visualized as a
rotation in the Bloch-sphere picture. The effect of unitary
conjugation is to apply the same operation on a transformed basis.
In the actual case, the matrix in Eq.~\eqref{eq:GKSpm} can be rewritten as
\begin{equation}
  \label{eq:gksrot}
  C_{\theta,\phi} =
  {\mathcal{R}}(\theta/2,\phi,\psi)
  \begin{pmatrix}
    0 & & \cr
     & 0 & \cr
     & & 1
  \end{pmatrix}
  {\mathcal{R}}^T(\theta/2,\phi,\psi).
\end{equation}
The matrix between the two rotations defines a phase damping about the
$z$ axis. Thus the process is a phase damping about an axis pointing
towards the spherical polar angles $\theta/2,\phi$. In fact the
rotation $U_2$ moves the $z$ axis to the direction described by the
spherical polar angles $\theta/2,\phi$, so the direction of the phase
damping is ``half way'' between the $z$ axis, and its transform by
$U_2$. Note that the angle $\psi$ is irrelevant in
Eq.~\eqref{eq:gksrot}, as it does not influence the polar angle of the
rotated $z$ axis. The phase damping channel has a single direction
which is special, thus it is necessarily described by two parameters.
In the case if $\hat U_2$ is traceless because $\theta=\pi+2k\pi$ is
satisfied, we obtain a phase damping channel about an axis in the $xy$
plane. We can conclude that the controlled-U gate with a traceless
$U_2$ is capable of simulating a generator of an arbitrary phase
damping channel in our scheme.

Returning to the generic GKS matrix in Eq.~\eqref{eq:gengks} we find
that $\rank C_{\theta,\phi,\psi}=1$, thus this general GKS matrix also
describes a phase damping channel, physically the same process as in
the traceless case. The only difference is that the evolution is now
accompanied by a coherent part, generated by the Hamiltonian in
Eq.~\eqref{eq:H1}.

\section{Control via teleportation}
\label{sect:telepcont}

\begin{figure}[htbp]
  \centering
    \includegraphics[width=0.4\textwidth]{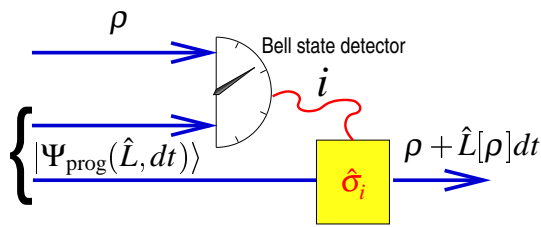}
  \caption{Bennett's teleportation scheme as a programmable circuit for Markovian decoherence}
  \label{fig:telepscheme}
\end{figure}

In this Section we briefly describe another simulation scheme based on
Bennett's quantum teleportation. It is depicted in
Fig.~\ref{fig:telepscheme}
The initial state $\varrho$ impinges at the input of a teleportation
arrangement. Two additional qubits are prepared in an entangled state
\begin{equation}
  \label{eq:progsttel}
  |\Psi_{\text{prog}}\rangle=
 \sqrt{1-\varepsilon}  |B_0\rangle+
  \sqrt{\varepsilon} \left(
\alpha _1  |B_1\rangle + \alpha _2  |B_2\rangle + \alpha _3
 |B_3\rangle  \right),
\end{equation}
with $|\alpha _1|^2+|\alpha _2|^2+|\alpha_3|^2=1$, which serves as the
\emph{program state}. We use the notation
\begin{eqnarray}
|B_0\rangle=\frac{1}{\sqrt{2}}
       \left( |0\rangle|1\rangle - |1\rangle|0\rangle \right)\; ; \nonumber \\
|B_1\rangle=\frac{1}{\sqrt{2}}
       \left( |0\rangle|0\rangle - |1\rangle|1\rangle \right)\; ; \nonumber \\
|B_2\rangle=\frac{1}{\sqrt{2}}
       \left( |0\rangle|0\rangle + |1\rangle|1\rangle \right)\; ;\nonumber \\
|B_3\rangle=\frac{1}{\sqrt{2}}
       \left( |0\rangle|1\rangle + |1\rangle|0\rangle \right).
\label{eq:bell}
\end{eqnarray}
for the elements of the Bell basis.  Then the usual Bennett
teleportation is carried out: a Bell state measurement on the two
appropriate qubits is carried out, and depending on the result, the
appropriate unitary transformation $\hat \sigma_i$ (the identity
operator or one of the Pauli operators) is carried out.

For $\varepsilon=0$ we have $ |\Psi_{\text{prog}}\rangle=|B_0\rangle$,
the state $\varrho$ is simply teleported: the identity operator is
implemented. However, for nonzero $\varepsilon$ we obtain
\begin{equation}
  \label{eq:rhoo}
  \varrho'= (1-\varepsilon) \varrho + \varepsilon \left(
    |\alpha_1|^2 \hat \sigma_1 \varrho \hat \sigma_1 +
    |\alpha_2|^2 \hat \sigma_2 \varrho \hat \sigma_2 +
    |\alpha_3|^2 \hat \sigma_3 \varrho \hat \sigma_3 \right)
\end{equation}
as a ``teleported'' state. Thus the scheme is essentially equivalent
to a random application of the Pauli operators. Setting
$dt=\varepsilon$ as in the previous Section, we can define
 \begin{equation}
   \label{eq:gentp}
   \hat L \varrho=    |\alpha_1|^2 \hat \sigma_1 \varrho \sigma_1 +
    |\alpha_2|^2 \hat \sigma_2 \varrho \sigma_2 +
    |\alpha_3|^2 \hat \sigma_3 \varrho \sigma_3 -
   \varrho.
 \end{equation}
A straightforward calculation shows that the corresponding GKS matrix
reads
\begin{equation}
  \label{eq:GKStel}
  {\mathcal C}(\alpha_1,\alpha_2,\alpha_3)=
  \begin{pmatrix}
|\alpha _1|^2 &                 &                   \cr
                             &  |\alpha _2|^2   &                   \cr
                             &                &     |\alpha _3|^2
  \end{pmatrix}.
\end{equation}
We have obtained a GKS matrix of rank 3, describing a generic Pauli
channel. In this case we have the parameters of the dynamics encoded
into the program state, too.

It is worth noting here that this second scheme is indeed
irreversible. However, the same process could be simulated in a
similar reversible framework as in Section~\ref{sec:progu}, utilizing
two ancillary qubits, and the universal programmable quantum gate
array in Ref.~\cite{fortschr49_987,pra65_022301}.

\section{Summary}
\label{sect:summary}

We have investigated quantum computational schemes for the simulation
of a generator of Markovian dynamics on a qubit, where the value of
the infinitesimal time step is encoded in a quantum state at the
input of the device. We have described the capabilities of the
arrangement if the ancilla to be applied is restricted to a single
qubit by characterizing the most general Lindbladian that can be
simulated. We have found that under these restrictions, a phase
damping about an arbitrary axis can be simulated. We have also
investigated a the Bennett quantum teleportation scheme as a possible
quantum circuit for performing a similar task.

The inevitable advantage of this method compared with a classical
simulation of quantum dynamics is that it works for any, even unknown
initial state, which may emerge as an output of another quantum
calculation.

The analysis can be carried out for systems ancillas of larger size
but in that case, the number of parameters grows, and the problem is
less transparent. We remark that for instance, the two-qubit program
state enables non-unital dynamics as well.

We believe that the study of simple quantum systems such as those
described here facilitates the understanding decoherence in general.

\acknowledgements This work was supported  by the European Union  projects QGATES and
CONQUEST,  by the Slovak Academy of Sciences via the project CE-PI,
by the project APVT-99-012304, and by the Research Fund of
Hungary (OTKA) under contract No. T034484.


\end{document}